\documentstyle[prl,aps,epsfig,twocolumn]{revtex}
\begin{document}
\draft
\flushbottom
\twocolumn[
\hsize\textwidth\columnwidth\hsize\csname @twocolumnfalse\endcsname

\title{Superconductivity in carbon nanotube ropes}
\author{J. Gonz\'alez   \\}
\address{
        Instituto de Estructura de la Materia.
        Consejo Superior de Investigaciones Cient{\'\i}ficas.
        Serrano 123, 28006 Madrid. Spain.}
%%        $^2$Instituto de Ciencia de Materiales.
%%        Consejo Superior de Investigaciones Cient{\'\i}ficas.
%%        Cantoblanco. 28049 Madrid. Spain.}

\date{\today}
\maketitle
\begin{abstract}
\widetext
We investigate the conditions in which superconductivity may
develop in ropes of carbon nanotubes. It is shown that the
interaction among a large number of metallic nanotubes favors
the appearance of a metallic phase in the ropes, intermediate
between respective phases with spin-density-wave and
superconducting correlations. These arise in samples with about 
100 metallic nanotubes or more, where the long-range Coulomb 
interaction is very effectively reduced and it may be overcome 
by the attractive interaction from the exchange of optical 
phonons within each nanotube. We estimate that the probability 
for the tunneling of Cooper pairs between neighboring nanotubes 
is much higher than that for single electrons in a disordered rope.
The effect of pair hopping is therefore what establishes the 
intertube coherence, and the tunneling amplitude of the Cooper 
pairs dictates the scale of the transition to the superconducting 
state.

\end{abstract}

]

\narrowtext 
\tightenlines

%\newpage

\section{Introduction}

During the last years there has been growing evidence of the 
existence of superconducting correlations in carbon nanotubes.
In the first place, the proximity effect has been observed 
in ropes of nanotubes placed between superconducting 
contacts\cite{kas,marcus}. Quite remarkably, it has been shown 
that the nanotubes may support supercurrents below the critical 
temperature of the contacts\cite{kas}. Later on, there have been 
experiments directed to probe the superconductivity inherent to
the nanotubes\cite{koc}. In some of the ropes, a transition has
been observed at a temperature below $1 \; {\rm K}$, with a drop 
of two orders of magnitude in the resistance down to a value 
consistent with the finite number of channels in the rope. 

From a theoretical point of view, the electronic interactions 
in the nanotubes have been also the subject of much 
attention\cite{bal,berk,eg,kane}. The nanotubes may show metallic 
properties depending on the helicity with which the graphene
sheet is wrapped to form the tubule. The experimental 
results\cite{wild} have agreed on that point with the 
theoretical predictions\cite{saito}. Given that the conduction 
takes place in a one-dimensional (1D) structure, it has been 
proposed that the nanotubes should be ideal systems for the 
observation of the so-called Luttinger liquid 
behavior\cite{emery,lutt1,lutt2,sch}. Actually, there have been
measurements providing evidence of the power-law dependence of 
the tunneling conductance\cite{exp,yao}, what is a signature of 
the mentioned behavior. These experiments seem therefore to
probe a regime in which the repulsive Coulomb interaction turns
out to be dominant in the nanotubes.

On the other hand, the interaction with the elastic modes of 
the nanotube plays an important role in the development of the 
superconducting correlations\cite{prl}. The magnitude of the 
supercurrents measured in Ref. \onlinecite{kas} is in some 
instances up to 40 times larger than expected from the estimate 
within the conventional proximity effect. It has been shown that 
the temperature dependence of the supercurrents is characteristic 
of the 1D behavior of the system. Moreover, their 
values can be only explained by taking into account the 
attractive electronic interaction coming from the exchange of 
phonons, on top of the repulsive Coulomb interaction\cite{prl}.

Recently, a microscopic model has been elaborated to account for 
the observation of superconductivity intrinsic to the ropes of 
nanotubes\cite{micro}. In these systems, the Coulomb interaction
can be significantly reduced depending on the number of metallic
nanotubes in the rope. One has to incorporate therefore the 
balance between the repulsive electron interaction and
the effective attractive interaction coming from phonon
exchange. It has been shown that, in the case of samples with
about 300 nanotubes like those displayed in Ref. 
\onlinecite{koc}, the superconducting correlations prevail in
the system. The intertube coherence is established mainly through 
the tunneling of Cooper pairs, which gives rise to the 
superconductivity in the bulk under suitable 
conditions\cite{micro}.

One of the aims of the present paper is to unveil the 
different phases that arise in the competition between the 
repulsive Coulomb interaction and the attractive interaction 
from phonon exchange in the ropes of nanotubes. For this 
purpose we will map the phase diagram of these systems taking 
the strength of the phonon couplings and the number of 
metallic nanotubes as the relevant variables. We will show 
that the region where the superconductivity may develop opens 
up for relatively large contents of metallic nanotubes. 
The phase with spin-density-wave correlations characteristic
of a repulsive interaction is confined to the cases where 
there are only a few nanotubes in the rope. 

We assume in any event that the formation of spin or charge 
order in the bulk is prevented by the fact that the nanotubes 
have a random distribution of helicities in the ropes. These 
do not have a crystalline structure from the three-dimensional
point of view, and the only possible long-range order arises 
in the superconducting phase. We will see that, between the phases 
with superconducting and spin-density-wave correlations, there is 
a metallic phase with no signal of instability in any 
direction. This is the phase to be found out most likely when 
measuring ropes with a number below $\sim 100$ 
metallic nanotubes.

We will also address the question of the maximum transition
temperature that can be reached in the ropes. There have
been recent experimental measures in carbon nanotubes of 
minimum diameter inserted in a zeolite matrix, from which a
critical temperature of about $15 \; {\rm K}$ has been 
inferred\cite{chi}. The small radius and high curvature of
those nanotubes have to lead to an enhancement of the 
electron-phonon coupling. This would explain in turn the
increase in the critical scale for superconductivity. 

One has to bear in mind, however, that the correlations
in individual nanotubes provide an indirect measure of 
the superconducting state. We will show that, in the 
three-dimensional structure of the rope, the transition
temperature is dictated by the amplitude for the Cooper 
pairs to tunnel between neighboring nanotubes. Even in
samples with large number of metallic nanotubes, the 
opening of the superconducting phase depends at last on
the existence of intertube coherence. The most efficient
way to increase the transition temperature may come actually
from devising some mechanism to enhance the tunneling rate
of the Cooper pairs between the nanotubes.

The content of this paper is organized as follows. In 
Section II we analyze the origin of the different 
interactions in the individual nanotubes. Section III is 
devoted to elaborate the model that takes into account 
the interaction among the metallic nanotubes, ending up 
with the discussion of the possible phases of a rope. 
In Section IV we incorporate the effect of the tunneling
of the Cooper pairs, performing some estimates of the 
transition temperature under different conditions.
Finally, Section V is devoted to discuss some of the 
assumptions under which our general analysis applies.

\section{Electron interactions in carbon nanotubes}

\begin{figure}
\begin{center}
\mbox{\epsfxsize 7cm \epsfbox{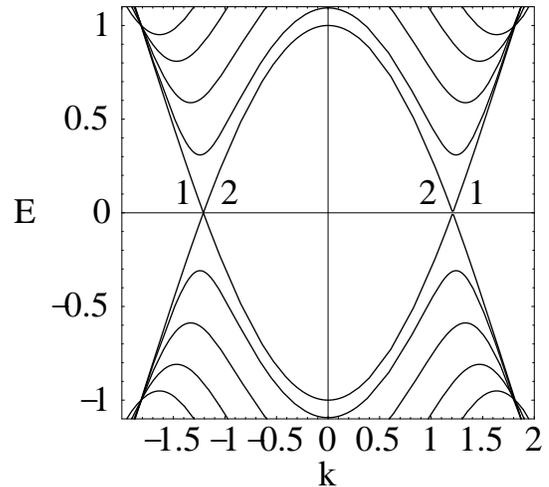}}
\end{center}
\caption{Different subbands of a (10,10) armchair nanotube.
The energy $E$ is measured in units of the hopping parameter
$t$ and the momentum $k$ is in units of the inverse of the 
lattice spacing.}
\label{metal10}
\end{figure}

We begin by considering the band structure of the individual
metallic nanotubes. We focus on the so-called armchair 
nanotubes, which have the structure depicted in Fig. 
\ref{long} of the Appendix A. They have gapless subbands 
irrespective of the influence of the nanotube curvature on
the orbitals. Fig. \ref{metal10} represents for instance 
the spectrum of a (10,10) armchair nanotube, which has a
radius $R \approx 15 a/ \pi $ in terms of the carbon-carbon
distance $a$. We will concentrate on the case of undoped
nanotubes, in which the Fermi level is placed at $E = 0$
in the scale of the figure.

The most important feature for the study of the
low-energy properties is the crossing of two subbands at the
Fermi level, that we label with the indices 1 and 2 as shown
in the figure. These subbands correspond to the modes with
vanishing transverse momentum in the nanotube, which are
governed by the one-particle hamiltonian\cite{dirac}
\begin{equation}
{\cal H} =    t  \left(
\begin{array}{cc}
0  &   1 - 2 \cos (\sqrt{3} k a/2)    \\
1 - 2 \cos (\sqrt{3} k a/2)    &   0
\end{array}         \right)
\end{equation}
$t$ being the hopping amplitude between neighboring sites.

The amplitudes of the modes behave differently in the 
nanotube lattice depending on whether they belong to one or 
the other subband. In the case of the subband with bonding 
character, the amplitude is a smooth function of the position, 
while the modes in the other subband have an amplitude that 
alternates the sign between the two sublattices of the 
honeycomb lattice. This property has important consequences 
regarding the form of the electron-electron interactions as 
well as the coupling of the electrons to the elastic modes 
of the lattice.

\begin{figure}
\begin{center}
\mbox{\epsfxsize 7cm \epsfbox{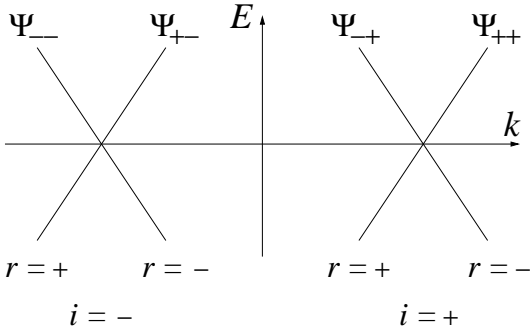}}
\end{center}
\caption{Linear branches in the low-energy spectrum of a 
carbon nanotube.}
\label{branch}
\end{figure}

In what follows we undertake the analysis of the low-energy
regime in which the physical properties are dominated by
the branches with approximate linear dispersion near the Fermi
level. For this purpose, we introduce an energy cutoff $E_c$ 
below the energy of the bottom of the first unoccupied
subband, together with its counterpart $-E_c$ below the Fermi 
level. The different branches are labelled by an index
$r = \pm$ denoting the right or left-moving character, and
by the Fermi point $i = \pm$ to which they are attached, as
shown in Fig. \ref{branch}.
Correspondingly, we end up with a collection of electron
fields $\Psi_{r i \sigma} (x)$, to which one more index $a$ 
will be added later on to label the nanotube in the rope.

The electron interactions that we are going to consider
have in general the form
\begin{eqnarray}
   \int_{-k_c}^{k_c} dk   \int_{-k_c}^{k_c} dp
     \int_{-k_c}^{k_c} dq
              \;\;\;\;\;\;\;\;\;\;\;\;\;\;\;
    \;\;\;\;\; \;\;\;\;\;\;\;\;\;\;&   &      \nonumber       \\
   \Psi_{r i \sigma}^{\dagger}(p+k) \Psi_{s j \sigma }(p)  V(k)
  \Psi_{t k \sigma '}^{\dagger}(q-k) \Psi_{u l \sigma '}(q)   &  &
\end{eqnarray}
The processes between the different branches can be
cataloged according to the notation of Ref.
\onlinecite{berk}. They fall into 12 different classes with
respective couplings denoted by $g_i^{(j)}$, the indices taking
the values $i = 1,2,4$ and $j = 1, \ldots 4$. The lower label
pays attention to the Fermi points of the respective fields,
following the convention that assigns $i = 1$ to backscattering
processes, $i = 2$ to scattering between currents at
different Fermi points, and $i = 4$ to scattering
between currents at the same Fermi point. The upper label
follows the same rule classifying the different combinations
of left-movers and right-movers, including the possibility of
having Umklapp processes ($j = 3$). 

We analyze next the contributions to the different scattering
processes from the interactions present in the nanotubes.

{\bf Electron-electron interactions.}

\begin{figure}
\begin{center}
\mbox{\epsfxsize 7cm \epsfbox{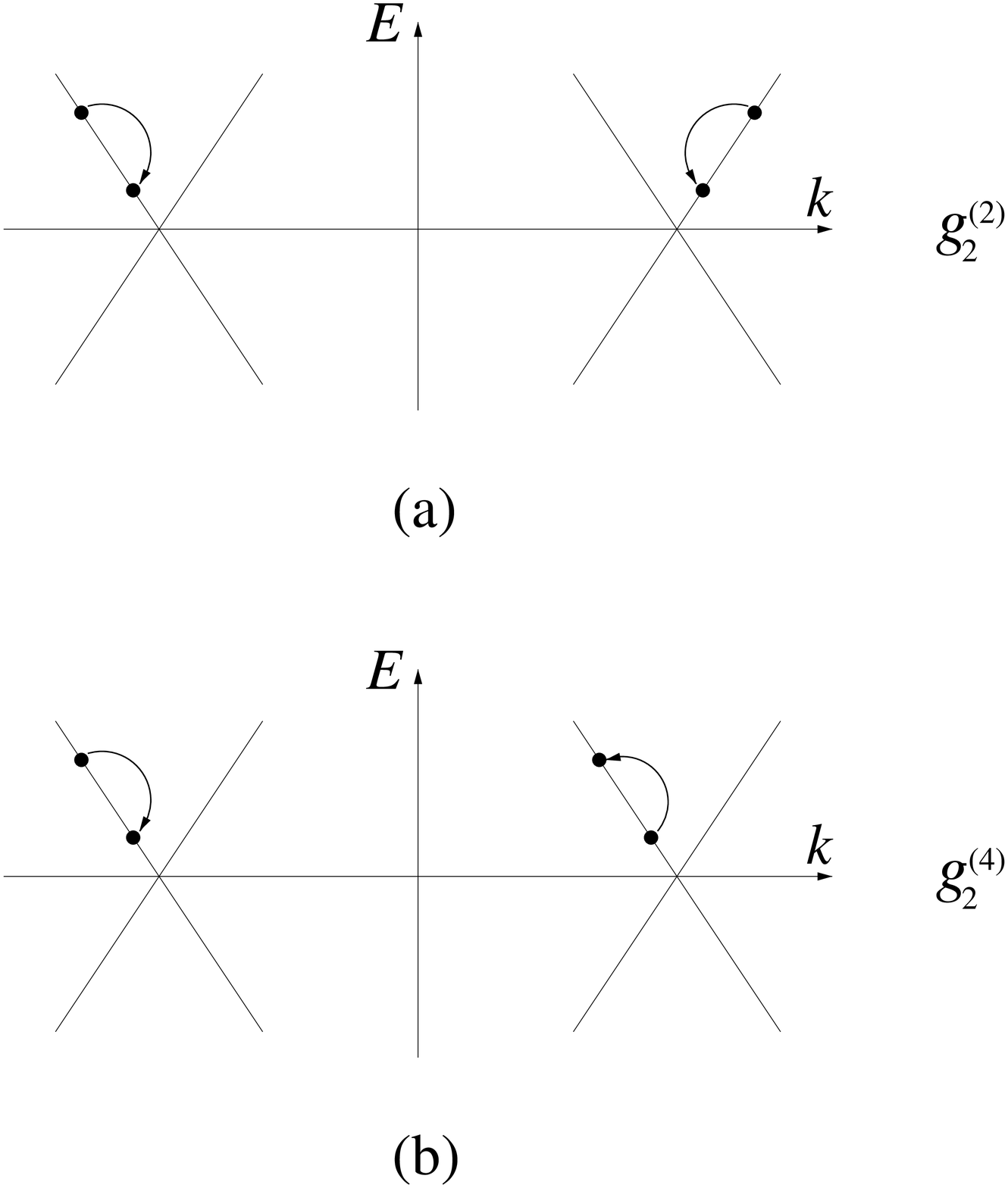}}
\end{center}
\caption{Small momentum-transfer processes which are enhanced
by the long-range Coulomb interaction.}
\label{geol2}
\end{figure}

\begin{figure}
\begin{center}
\mbox{\epsfxsize 7cm \epsfbox{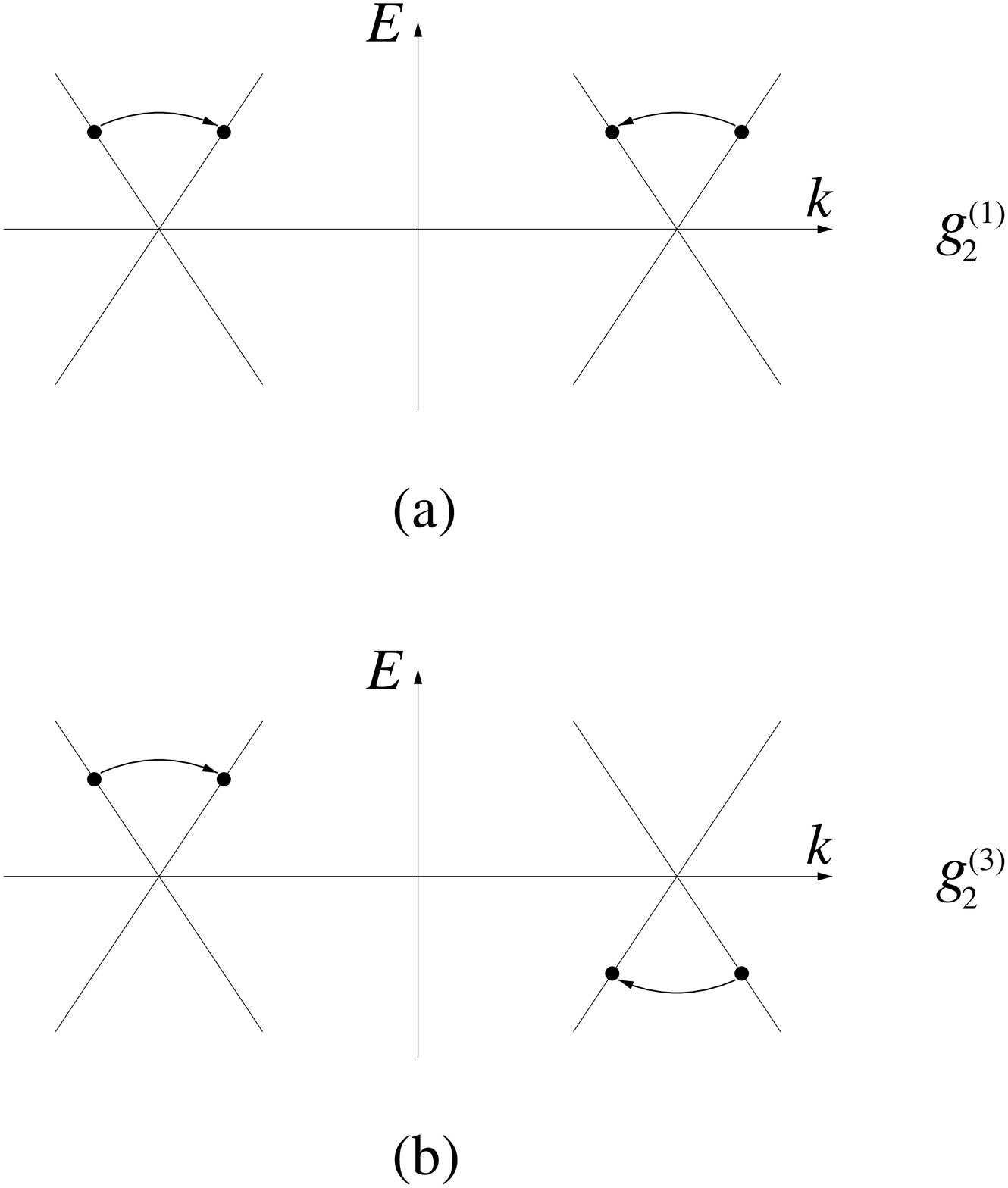}}
\end{center}
\caption{Flavor-changing processes with small contribution from
the Coulomb interaction.}
\label{geol}
\end{figure}

We deal with the situation in which the Coulomb interaction
is not screened by external gates. It is known that the 
Coulomb potential remains long-ranged in one spatial 
dimension\cite{grab}. The appropriate expression in momentum 
space is\cite{wang}
\begin{equation}
V_C (k) = (e^2 /4\pi^2) \log |(k_c + k)/ k|
\end{equation}
The repulsive interaction is therefore enhanced at 
small momentum transfer, as in the processes shown in 
Fig. \ref{geol2}. 
However, processes like those in Fig. \ref{geol}, in which the
flavor $r$ is changed by the interaction, are highly suppressed.
This is due to the fact that, as mentioned before, the modes
corresponding to one of the gapless subbands alternate the sign
in the two sublattices of the nanotube. When the distance
between the two currents that interact is much larger than the
nanotube radius, the matrix element of the interaction averages
to zero over the section of the nanotube.

Thus, the long-range repulsive interaction reflects in an 
enhancement of the interbranch and intrabranch couplings 
$g_2^{(2)}$, $g_2^{(4)}$, $g_4^{(2)}$ and $g_4^{(4)}$.
The flavor-changing processes associated to $g_2^{(1)}$,
$g_2^{(3)}$, $g_4^{(1)}$ and $g_4^{(3)}$ survive in the form 
of a residual short-range interaction. Its strength, however, 
is corrected by a factor of the order of $\sim 0.1 \; a/R$ in 
comparison to the nominal strength of the long-range Coulomb 
interaction\cite{eg,kane}. This represents in general a 
relative reduction by two orders of magnitude, for the 
nanotubes that are typically found in the ropes.

There remain the processes with momentum transfer of the order
of $\sim 2 k_F$. These processes probe the interaction at short 
distances. It has been shown that the corresponding couplings 
are also reduced by a factor of the order of $\sim 0.1 \; a/ R$ 
relative to the strength $e^2 /4 \pi^2 $ of the 
forward-scattering processes\cite{eg,kane}.

The couplings for large momentum transfer, as well as 
$g_2^{(1)}$, $g_2^{(3)}$, $g_4^{(1)}$ and $g_4^{(3)}$, are 
marginally relevant in the renormalization group sense. 
This means that they have greater strength as the theory is
scaled to smaller energies, but the rate of increase starts
being quadratic in the own couplings\cite{berk,eg}. 
Thus, one has to proceed
to extremely low energies, many orders of magnitude below
$E_c$, to have a significant increase of the couplings with
respect to their nominal values\cite{eg}.

{\bf Electron-phonon interactions.}

The electron-phonon interactions give rise to a retarded
interaction between the electrons, which may become attractive
at frequencies below the typical energy of the phonons. 
The potential for the effective interaction has the form
\begin{equation}
V (k,\omega ) = - g_{p,p'}(k) g_{q,q'}(k) 
  \frac{\omega_k }{- \omega^2 + \omega_k^2 }
\end{equation}
where $\omega_k $ is the phonon energy and the $g_{p,p'}(k)$ 
are appropriate electron-phonon couplings described in the
Appendix A. The form of the interaction is 
dictated by the latter, whose dependence on the momentum 
transfer varies according to the kind of phonons exchanged.

{\em Acoustic phonons.}--- 
In this case,  $\omega_k $ is proportional to
$|k|$ at small momentum transfer. This means that, regarding the 
interbranch and intrabranch processes, the range of energies in
which the effective interaction becomes attractive is greatly
reduced. This instance has been studied with detail in Ref.
\onlinecite{loss}, looking for the regime in which the 
retarded interaction may drive the transition to a state with
superconducting correlations. Taking the parameters
appropriate for the nanotubes, it can be
seen that the critical exponents estimated in this way are
very small. Therefore, the acoustic phonons at small momentum 
transfer do not play a role in the superconductivity
observed in the nanotubes.
 
Moving now to the processes with large momentum transfer 
$k \sim 2 k_F$, the phonon energy $\omega_{2 k_F}$ becomes 
of the order of the Debye frequency $\omega_D$. This provides
then a significant range in which the effective interaction 
becomes attractive. The precise couplings to the acoustic
phonons in the nanotubes can be found in Ref. \onlinecite{jdd}.
It has been shown there that some restrictions arise in the 
exchange of phonons depending on the bonding or antibonding
character of the subbands involved. In the case of longitudinal 
acoustic phonons, the processes cannot take one electron from a 
given subband to the other. Moreover, the couplings have 
opposite signs when the scattered electron is in one or the 
other subband. The exchange of phonons leads then to the 
effective couplings below the Debye frequency
\begin{equation}
g_{{\rm eff},1}^{(1)} = - g_{{\rm eff},1}^{(3)} < 0
\label{klarge}
\end{equation}

In the case of the transverse acoustic phonons, the 
phonon-exchange processes in which the electrons remain in the 
same subband are forbidden. Processes associated to 
$g_{1}^{(2)}$ and $g_{1}^{(4)}$ in which 
the electron shifts from one subband to the other could play 
in principle a role, but there is a kinematic factor that 
vanishes when the incoming and outgoing modes have opposite
momenta. There are therefore no more contributions from the
acoustic phonons when the system is precisely at half-filling.

{\em Optical phonons.}--- The Debye frequency is of the 
same order that the energy of the acoustic phonons at $2 k_F$, 
so that we can expect the effective couplings to be comparable 
to those in Eq. (\ref{klarge}). 
The different symmetry of the modes with bonding and  
antibonding character imposes again restrictions in the way
the optical phonons can be exchanged, as shown in the 
Appendix A. In the case of the transverse optical phonons,
the electron modes have to remain within the same subband,
while the longitudinal phonons force the electron to shift
from one of the subbands to the other. The effective
couplings that we get have the following signs
\begin{eqnarray}
g_{{\rm eff},2}^{(1)}  & = &  - g_{{\rm eff},2}^{(3)} < 0    \\
g_{{\rm eff},2}^{(2)}  & = &  - g_{{\rm eff},2}^{(4)} < 0    \\
g_{{\rm eff},4}^{(1)}  & = &  - g_{{\rm eff},4}^{(3)} < 0    \\
g_{{\rm eff},4}^{(4)}  & = &  - g_{{\rm eff},4}^{(2)} < 0   
\end{eqnarray}

We finally remark that the Debye frequency of the phonons 
in the nanotubes can be estimated to be between 0.1 and 
$0.2 \; {\rm eV}$. This is of the same order of magnitude that 
the cutoff $E_c$, that marks the energy below which
the nanotubes can be viewed as simple 1D systems. In that
regime, the effective interactions coming from the exchange of
phonons can be matched against the interaction of purely
electronic origin. We will adopt henceforth this procedure, 
as long as the electronic properties in which we are going to 
be interested refer to energies much lower than $E_c$.

\section{Electronic interactions in ropes of nanotubes}

We turn now to the processes that take place between the 
nanotubes in a rope. A property to be taken into 
account is that the rope is made of a disordered mixture of 
nanotubes with all kind of diameters and  
helicities. Given their random distribution, the lattices of 
neighboring nanotubes are not aligned in general. The 
consequences of this fact have been studied in Ref.
\onlinecite{mkm}. As long as there is a mismatch in the position
of the respective Fermi points of neighboring nanotubes, the 
conservation of the longitudinal momentum cannot be precisely
enforced in the tunneling processes. At the one-particle level,
the electronic states appear to be localized along the individual
nanotubes, and the intertube coherence is strongly
suppressed.

The authors of Ref. \onlinecite{mkm} have computed the
expression of the tunneling amplitude between nanotubes in
the so-called {\em compositionally disordered} ropes.
In the case of nanotubes with the same helicity and aligned 
lattices, the estimate for the tunneling amplitude is
$t_T \sim 0.01 \; {\rm eV}$. When there is a mismatch $\delta k$
between the Fermi points of neighboring nanotubes, it has been 
found that the amplitude for an electron hopping between them
is suppressed by a relative factor $\sim \exp [-R a_0
(\delta k)^2 /4]$, where $a_0$ is a parameter of the order of 
$\sim 0.5$ \AA. Taking a    
typical nanotube radius $R = 7$ \AA, the average value of that
factor is of the order of $\sim 0.005$.

The presence of compositional disordered has also important 
consequences regarding the kind of electronic instabilities that
may develop in the rope. Depending on the point of the phase   
diagram, the individual nanotubes may support $2 k_F$
charge-density-wave and superconducting correlations, or 
$2 k_F$ spin-density-wave correlations. It is clear, however, 
that the absence of a three-dimensional crystalline structure 
precludes the
development of charge or spin order in the bulk of the rope.
The transition to a superconducting state is therefore favored
when the correlations are strong enough to force the symmetry
breaking in that direction.

The possibility of having a superconducting transition in the 
rope of nanotubes arises from the balance of the effective 
attractive interactions due to phonon exchange and the  
repulsive interactions of purely electronic character. At the
diagrammatic level, for instance, the correlations with $s$-wave 
order parameter are driven by the couplings $g_2^{(2)}$, 
$g_1^{(1)}$, $g_2^{(1)}$ and $g_1^{(2)}$, as shown in the 
Appendix B \cite{caron}. The obstruction to
having superconducting correlations in the nanotubes
comes, in principle, from the large contribution of the Coulomb
potential to the coupling $g_2^{(2)}$.
In this respect, the processes associated to $g_1^{(1)}$ and
$g_2^{(1)}$ contribute to enhance the superconducting
correlations when they are able to develop, but they play a
secondary role in the determination of the phase diagram of the 
ropes.

It is therefore necessary to build a model that takes into
account the competition between attractive and repulsive
interactions in the interbranch and intrabranch processes
$g_2^{(2)}$, $g_2^{(4)}$, $g_4^{(2)}$ and $g_4^{(4)}$.
This model can be solved by means of bosonization   
techniques. With this method, one boson field is introduced for
each of the linear branches below $E_c$ in each metallic
nanotube. At this point we need
to introduce a new index $a$ labelling the metallic nanotubes
in the rope, so that the fermion fields read now
$\Psi^{(a)}_{r i \sigma }(x)$. The correspondence between each of
these and the respective boson operator is given by 
\begin{equation}
: \Psi^{(a) \dagger}_{r i \sigma}(x) \Psi^{(a)}_{r i \sigma }(x) : =
   \rho^{(a)}_{r i \sigma }(x)
\end{equation}
where the colons represent normal ordering and
$\rho^{(a)}_{r i \sigma }(x)$ stands for the density operator
associated to the given branch.

Neglecting for the time being the tunneling between the
nanotubes, the hamiltonian for the model, including a collection
$a = 1, \ldots n$ of metallic nanotubes, takes the form   
\begin{eqnarray}
H_0  & = &    \frac{1}{2} v_F \int_{-k_c}^{k_c} dk
 \sum_{a r i \sigma }  : \rho^{(a)}_{r i \sigma} (k)
             \rho^{(a)}_{r i \sigma} (-k)  :   \nonumber      \\   
  &  &    + \frac{1}{2}  \int_{-k_c}^{k_c} dk \;
      \sum_{a r i \sigma } \rho^{(a)}_{r i \sigma} (k) \;  
  \sum_{b s j \sigma'  }  V^{(ab)}_{r i, s j} (k)  \;
                      \rho^{(b)}_{s j \sigma'} (-k) 
\label{ham}
\end{eqnarray}
where $k_c$ is related to $E_c$ through the Fermi velocity
$v_F$, $k_c = E_c/v_F$.
   
The interaction term in (\ref{ham}) can be better organized
by introducing the charge density operators
\begin{equation}
\rho^{(a)}_{r i \rho }(x)  = \frac{1}{\sqrt{2}} 
 \left(  \rho^{(a)}_{r i \uparrow }(x) +
       \rho^{(a)}_{r i \downarrow }(x)   \right)
\end{equation}
and their spin density counterparts. We define
further the linear combinations
\begin{eqnarray}
\tilde{\rho}^{(a)}_{1 \rho }(x) & = &
                  \rho^{(a)}_{+ + \rho }(x) +
              \rho^{(a)}_{- - \rho }(x)                \\
\tilde{\rho}^{(a)}_{2 \rho }(x) & = &
                   \rho^{(a)}_{+ - \rho }(x) +
              \rho^{(a)}_{- + \rho }(x)
\end{eqnarray}
Obviously, $\tilde{\rho}^{(a)}_{1 \rho }(x)$ is made of the
modes with bonding character, while
$\tilde{\rho}^{(a)}_{2 \rho }(x)$ contains those
with antibonding character.

We recall that, in the case of the attractive interaction from
the exchange of phonons, the sign of the coupling depends on the
bonding or antibonding character of the currents involved. Thus,
a great simplification is achieved by introducing the operators
\begin{eqnarray}
\tilde{\rho}^{(a)}_{+ \rho }(x) & = &   \frac{1}{\sqrt{2}}
   \left(   \tilde{\rho}^{(a)}_{1 \rho }(x) +
              \tilde{\rho}^{(a)}_{2 \rho }(x)   \right)     \\  
\tilde{\rho}^{(a)}_{- \rho }(x) & = &  \frac{1}{\sqrt{2}}
    \left(    \tilde{\rho}^{(a)}_{1 \rho }(x) -
              \tilde{\rho }^{(a)}_{2 \rho }(x)   \right)
\end{eqnarray}
The Coulomb interaction acts on the symmetric combinations
$\tilde{\rho}^{(a)}_{+ \rho }(x)$, while the effective attractive
interaction only couples the antisymmetric combinations
$\tilde{\rho}^{(a)}_{- \rho }(x)$. The hamiltonian (\ref{ham})
can be rewritten in the form
\begin{eqnarray}
H_0  & = &    \frac{1}{2} v_F \int_{-k_c}^{k_c} dk
 \sum_{a r i \sigma }  : \rho^{(a)}_{r i \sigma} (k)
             \rho^{(a)}_{r i \sigma} (-k)  :   \nonumber    \\   
  &  &    + \frac{1}{2}  \int_{-k_c}^{k_c} dk \;
    \left(  \sum_a \tilde{\rho}^{(a)}_{+ \rho} (k) \;
         V_C (k)   \sum_b     \;
   \tilde{\rho}^{(b)}_{+\rho} (-k)   \right.     \nonumber     \\
  &  &  \left.    \;\;\;\;\;\;\;\;\;\;\;\;\;\;\;\;\;\;
         +   g  \sum_a \tilde{\rho}^{(a)}_{- \rho} (k) \;
                \tilde{\rho}^{(a)}_{- \rho} (-k)    \right)
\label{ham2}
\end{eqnarray}  
where we have defined
\begin{equation}
g = g_2^{(2)} = g_4^{(4)} = - g_2^{(4)} = - g_4^{(2)}  
\end{equation}

After the decoupling of the Coulomb and the effective
attractive interaction, we observe that the latter affects to 
a number $n$ of the $4n$ density operators, while the Coulomb 
interaction acts only on the channel of the total charge density. 
That is, the hamiltonian (\ref{ham2}) can be completely 
diagonalized by changing variables to the symmetric combination 
of the charge density operators of all the metallic nanotubes. 
This is a consequence of the long-range character of the
Coulomb interaction, that allows us to assume that, in the 
interbranch and intrabranch processes,
the different nanotubes interact between them with the
same strength. 

The balance between the long-range repulsive interaction and
the effective attractive interaction is possible as the 
former turns out to be greatly reduced for large values of $n$.
The attraction from phonon exchange is
otherwise an intratube effect that does not depend on the number
of nanotubes. It is then interesting to draw the phase diagram
of the ropes in terms of the number of metallic nanotubes and
the strength of the attractive interaction.
At large values of $n$ and sufficiently large values of $|g|$,
there must exist a phase where the attractive interaction
prevails, with the development of superconducting correlations
in the individual nanotubes. For low values of $n$ and small $|g|$,
the electronic properties have to be dictated by the Coulomb
interaction, with the appearance of the spin-density-wave
correlations characteristic of the Luttinger liquid behavior.

The different phases can be identified by looking at the 
correlators of the model.
The bosonization techniques allow to compute
them by using the correspondence between the fermion fields
and the respective boson operators\cite{emery,lutt1}
\begin{eqnarray}
 \Psi^{(a)}_{r + \sigma }(x)  & = &
   \exp \left( i \Phi^{(a)}_{r + \sigma } (x) \right)      \\
 \Psi^{(a)}_{r - \sigma }(x)  & = &
   \exp \left( - i \Phi^{(a)}_{r - \sigma } (x) \right)   
\end{eqnarray}
where $\partial_x \Phi^{(a)}_{r i \sigma } (x) = 
 2\pi  \rho^{(a)}_{r i \sigma } (x)$ .

We deal first with the propagator $D_{\rm sc}^{(0)} (x,t)$
of the Cooper
pairs in the individual nanotubes. This object factorizes into 
the different channels that appear after diagonalizing the 
hamiltonian
\begin{eqnarray}
 D_{\rm sc}^{(0)}(x,t) & \equiv &
          \langle \Psi^{(a) \dagger}_{+ + \uparrow} (x,t)
  \Psi^{(a) \dagger}_{- - \downarrow} (x,t)
      \Psi^{(a)}_{- - \downarrow} (0,0)  
          \Psi^{(a)}_{+ + \uparrow} (0,0)  \rangle  \nonumber  \\
      &  =  &   C_{\rm sc}(x,t) \prod_{1}^{n} N_{\rm sc}(x,t)
               \prod_{1}^{3n-1} F (x,t)
\label{coop}
\end{eqnarray}
The first factor corresponds to the channel of the total charge 
density of the metallic nanotubes, while $N_{{\rm sc}}$ stands for 
the contribution of the antisymmetric combination of the charge in 
the two subbands. The rest of the factors corresponds to the 
channels with no interaction.

The different factors in Eq. (\ref{coop}) can be computed in 
terms of the respective boson propagators, what leads to 
expressions of the form
\begin{eqnarray}
X_{\rm sc}(x,t) & = &  \exp \left( -\frac{1}{2n} \int_{0}^{k_c}
 dk \;\;\;\;\;\;\;\;\;\;\;\;\;\;\;\;\;\;   \right.   \nonumber    \\
   &  &  \;\;\;\;\;\;\;\;\;  \left.  \frac{1}  {\mu (k)\: k}
   \left(1 - \cos (kx) \: \cos (\tilde{v}_F kt) \right) \right)
\label{prop}
\end{eqnarray}
In the case $X_{\rm sc}(x,t) = C_{\rm sc}(x,t)$, we have 
$\mu (k) = 1/\sqrt{1 + 8n V_{C}(k)/v_F}$ and $\tilde{v}_F (k) = 
v_F/\mu (k)$. For $X_{\rm sc}(x,t) = N_{\rm sc}(x,t)$, we have 
the constant parameters $\mu  = 1/\sqrt{1 - 4|g|/\pi v_F}$ and 
$\tilde{v}_F  = v_F/\mu $.

The same technique can be applied to the computation of the 
propagator $D_{\rm sdw}^{(0)}(x,t)$ of the spin-density-waves
along the nanotubes. The correlator is given by
\begin{eqnarray}
 D_{\rm sdw}^{(0)}(x,t) & \equiv &
          \langle \Psi^{(a) \dagger}_{+ + \uparrow} (x,t)
      \Psi^{(a)}_{- - \downarrow} (x,t)
         \Psi^{(a) \dagger}_{- - \downarrow} (0,0)
   \Psi^{(a)}_{+ + \uparrow} (0,0)  \rangle  \nonumber  \\
      &  =  &   C_{\rm sdw}(x,t) \prod_{1}^{n} N_{\rm sdw}(x,t)
               \prod_{1}^{3n-1} F (x,t)
\label{sdw}
\end{eqnarray}
Each of the factors in Eq. (\ref{sdw}) has now the 
representation
\begin{eqnarray}
X_{\rm sdw}(x,t) & = &  \exp \left( -\frac{1}{2n} \int_{0}^{k_c}
 dk \;\;\;\;\;\;\;\;\;\;\;\;\;\;\;\;\;\;   \right.   \nonumber    \\
   &  &  \;\;\;\;\;\;\;\;\;  \left.  \frac{ \mu (k) }  {k}
   \left(1 - \cos (kx) \: \cos (\tilde{v}_F kt) \right) \right)
\label{prop2}
\end{eqnarray}
with the same correspondence between $\mu (k)$ and $C$, $N$ and
$F$ as for the propagator of the Cooper pairs.

The correlators in the individual nanotubes cannot display an 
instability for any finite value of the frequency\cite{sch}.
The way to
discern the tendency to the formation of long-range order is
to determine whether the correlations are enhanced at large 
distances over the values in the absence of interaction. When
this happens, the Fourier transform of the corresponding
propagator at zero temperature diverges at zero frequency
and momentum. At nonvanishing temperature, the propagators
remain finite but the intertube coupling may give rise to 
the breakdown of the symmetry if the correlations in the 
nanotubes are sufficiently enhanced. We will address this 
question in the next section. The computation of the propagators 
(\ref{coop}) and (\ref{sdw}) can be extended to the case of 
temperature $T \neq 0$, just by inserting the factor
$1 + 2/[\exp (\tilde{v}_F |k|/T) - 1]$ in the integrand of
expressions such as (\ref{prop}) and (\ref{prop2}) \cite{emery}.

At this point, we stick to the model at zero temperature and
map the regions with enhanced superconducting or  
spin-density-wave correlations. The propagators (\ref{coop}) 
and (\ref{sdw}) do not show a perfect scaling behavior at large 
distances, due to the momentum dependence of the Coulomb 
potential. We may take however an effective value $k_0$ of the 
momentum in the infrared to approximate the behavior of each
propagator by a power-law with a constant exponent.

In the case of $D_{\rm sc}^{(0)}(x,t)$, the decay at large
distances takes the form
\begin{equation}
D_{\rm sc}^{(0)}(x,0) \sim 1/x^{2+\gamma}
\end{equation}
where the anomalous scaling dimension is given by 
\begin{equation}
\gamma = \frac{1}{2n \mu (k_0)} + \frac{1}{2 \mu } 
            - \frac{1}{2} - \frac{1}{2n}
\label{asd}
\end{equation}
Similarly, $D_{\rm sdw}^{(0)}(x,t)$ has a large-distance 
behavior
\begin{equation}
D_{\rm sdw}^{(0)}(x,0 ) \sim 1/x^{2+\delta}
\end{equation}
with an anomalous exponent
\begin{equation}
\delta = \frac{\mu (k_0)}{2n} + \frac{\mu }{2}     
            - \frac{1}{2} - \frac{1}{2n}          
\end{equation}

\begin{figure}
\begin{center}
\mbox{\epsfxsize 8.5cm \epsfbox{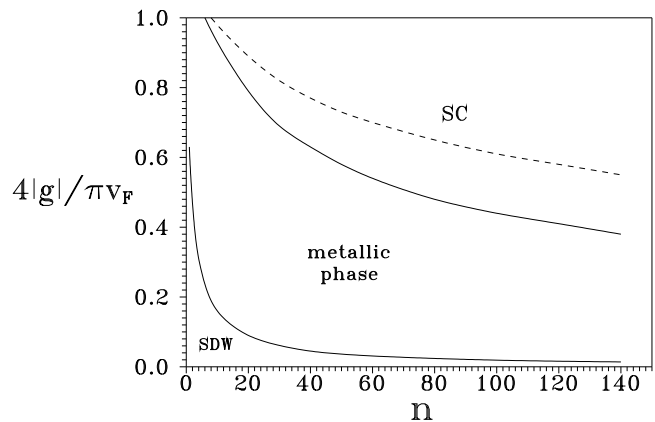}}
\end{center}
\caption{Phase diagram showing the regions where the 
superconducting (SC) correlations and the spin-density-wave
(SDW) correlations prevail, in terms of the strength $|g|$ of
the effective attractive interaction and the number $n$ of
metallic nanotubes.}
\label{scsdw}
\end{figure}

We have represented in Fig. \ref{scsdw} the different
phases that arise by varying the number $n$ of metallic 
nanotubes and the strength $|g|$ of the intratube attractive 
interaction. For the Coulomb interaction, we have taken the
values $2 e^2/\pi^2 v_F = 1.0$ and $k_0 = 10^{-3} k_c $,
which are appropriate for typical experimental 
samples\cite{prl}. At sufficiently large values of $n$, 
a phase with superconducting
correlations opens up, characterized by negative values of 
the anomalous dimension $\gamma $. This phase is placed 
above the upper full line in the diagram. Another phase with 
spin-density-wave correlations shows up, below the lower 
full line in the diagram. This region corresponds to the
points characterized by negative values of $\delta $.

We observe also the existence of an intermediate region in 
which neither $\gamma $ nor $\delta $ are negative. This 
represents a phase with intrinsic metallic properties, where 
no correlations are enhanced and no order would develop 
irrespective of the value of the intertube coupling in the 
rope\cite{foot}. In our model, this phase arises from the 
balance between the Coulomb and the effective attractive 
interaction. The former remains long-ranged, but the 
interaction among a large number of metallic nanotubes is 
what favors the appearance of the metallic phase, 
as a bridge to the development of the superconducting 
correlations.

\section{Superconducting transition in ropes of nanotubes}

In the preceding section we have neglected the effect of 
tunneling between metallic nanotubes, relying on the smallness 
of the amplitude for that process. We have seen however that, 
in the absence of intertube hopping, the model cannot develop 
any instability at finite temperature. 
In this respect, the upper full line in Fig. \ref{scsdw} has to 
be considered as a first approximation to the boundary where the
superconducting phase opens up in real samples.
The tunneling amplitude 
between neighboring nanotubes in a rope is actually what 
dictates the transition temperature to the superconducting 
state, as we study in what follows.

As mentioned before, the single-particle hopping between 
neighboring nanotubes is strongly suppressed in a disordered 
rope. The tunneling amplitude between nanotubes with different 
helicities can be estimated to be of the order of $\sim 0.5 
\times 10^{-4} \; {\rm eV}$. Its smallness stems from the 
obstruction to having precise momentum conservation in the 
hopping between the misaligned lattices of the 
nanotubes\cite{mkm}. In these conditions, the tunneling of 
Cooper pairs turns out to be a much more important effect.

The probabilities of the single-particle hopping and the pair 
hopping can be compared by measuring the respective amplitudes 
with respect to the energy cutoff $E_c$. We can take this 
quantity of the order of $\sim 0.1 \; {\rm eV}$, which is the 
scale below which the  nanotubes in a rope are seen as 1D 
objects. Thus, we find that the single-particle hopping has a 
relative weight of the order of $\sim 0.5 \times 10^{-3}$. The 
amplitude for the tunneling of Cooper pairs can be computed 
roughly as the square of the amplitude $t_T \sim 0.01 \; 
{\rm eV}$, so that the relative weight $\lambda_2$ of that 
process becomes of the order of $\sim 10^{-2}$.

The model of the preceding section has to be corrected then
by adding to the hamiltonian (\ref{ham2}) the term describing
the tunneling of the Cooper pairs
\begin{eqnarray}
 H_2 =  \sum_{\langle a,b \rangle}  ( \lambda_2 )_{ab}
  \int_{-k_c}^{k_c} dk   \int_{-k_c}^{k_c} dp
     \int_{-k_c}^{k_c} dp'
              \;\;\;\;\;\;\;\;\;\;\;\;\;\;\;
                  \;\;\;\;\;&   &      \nonumber       \\
   \Psi^{(a) \dagger}_{r i \uparrow}(k+p)
          \Psi^{(a) \dagger}_{-r -i \downarrow}(-p)
        \Psi^{(b)}_{-s -j \downarrow}(-p')
           \Psi^{(b)}_{s j \uparrow}(k+p')  &  &
\label{ham3}
\end{eqnarray}
where the sum runs over all pairs $\langle a,b \rangle$ of
nearest-neighbor metallic nanotubes.

Given that the model without pair hopping is exactly solvable,
we can compute the scaling dimension of the operator in $H_2$.
We can check in this way whether this is a relevant perturbation
or otherwise it fades away at low energies. The na\"{\i}ve 
dimension of the four-fermion operator is 2, and the deviation
from that value can be obtained from the large-distance behavior 
of the correlator in Eq. (\ref{coop}). The anomalous scaling 
dimension coincides with the exponent given in Eq. (\ref{asd}). 
Therefore, we observe that it is negative in the phase with 
superconducting correlations, what points at the enhancement of 
the pair-hopping perturbation at low energies. 

The values that can be obtained for $\gamma $ are rather small
anyhow. Taking again an effective value of $k_0 = 10^{-3} k_c$ , 
we find $\gamma \approx -0.02$ for $n = 60$ and 
$4|g|/\pi v_F = 0.6$, and
$\gamma \approx -0.16$ for $n = 120$ and $4|g|/\pi v_F = 0.8$.
In the latter case, for instance, the anomalous scaling leads to
an enhancement by a factor that, four orders of magnitude below
$E_c$, can be at most of the order of $(10^{-4})^{\gamma } \sim 4$. 
Such a scaling does not affect our estimate of the order of 
magnitude for the relative weight $\lambda_2 $.
  
\begin{figure}
\begin{center}
\mbox{\epsfxsize 8cm \epsfbox{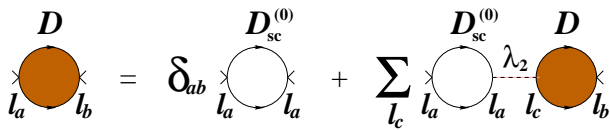}}
\end{center}
\caption{Self-consistent diagrammatic equation for the
propagator $D$ of Cooper pairs along the rope.}
\label{dyson}
\end{figure}

As long as we have the small parameter $\lambda_2 $ in the 
system, we may include the effects of pair hopping by summing
up all the processes in which a Cooper pair, propagating along 
a nanotube with the amplitude (\ref{coop}), tunnels to a 
neighboring metallic nanotube. Let us denote by $l_a$ the 
position of nanotube $a$ in the transverse section of the rope.
We will assume that the metallic nanotubes are dense in the 
collection of nanotubes of the rope\cite{priv}. 
Then, we may write the propagator of the Cooper pairs between 
the metallic nanotube $a$ and the metallic 
nanotube $b$ as a function $D (l_a,l_b;x,t)$. This object is
related to $D_{\rm sc}^{(0)}(x,t)$ through the self-consistent
equation represented in Fig. \ref{dyson}, in the approach that
takes $H_2$ as a perturbation to the hamiltonian 
(\ref{ham2}).

For the sake of simplifying the calculation, we will suppose
that the positions $l_a$ form a periodic arrangement in the
transverse section of the rope. Then we can take the Fourier
transform of these variables as well as of the distance $x$
along the nanotube. The equation for the Fourier transformed
propagator $\widetilde{D} (q;k,\omega_k)$ reads
\begin{equation}
\widetilde{D} (q;k,\omega_k) =
         \widetilde{D}_{\rm sc}^{(0)} (k,\omega_k) +
    \widetilde{D}_{\rm sc}^{(0)} (k,\omega_k) \lambda_2 (q)
                       \widetilde{D} (q;k,\omega_k)
\label{fourier}
\end{equation}

The measure of the condensation of Cooper pairs in the rope
is given by the propagator at zero frequency and momentum
$\widetilde{D} (0;0,0)$. This accounts for the propagation of
a Cooper pair from a metallic nanotube to the rest in the rope. 
From Eq. (\ref{fourier}) we obtain the expression
\begin{equation}
\widetilde{D} (0;0,0) = 
     \frac{ \widetilde{D}_{\rm sc}^{(0)} (0,0) }
  {  1 - \lambda_2 (0) \widetilde{D}_{\rm sc}^{(0)} (0,0)  }
\end{equation}

\begin{figure}
\begin{center}
\mbox{\epsfxsize 8cm \epsfbox{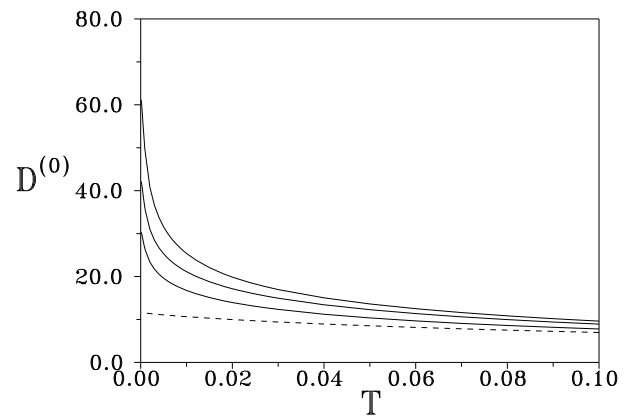}}
\end{center}
\caption{Plot of the propagator
$\widetilde{D}_{\rm sc}^{(0)}$ 
at zero frequency and momentum versus $T/E_c$, for
$2 e^2/(\pi^2 v_F) = 1.0$ . The dashed line
corresponds to the case $n = 1$ and $g = 0$, and the
solid lines to $n = 100$ and the respective values  
(from top to bottom) $4|g|/(\pi v_F) = 0.75, 0.5, 0$ .}
\label{corre}
\end{figure}

According to the conventional interpretation, the transition
to the superconducting state is given by the point at which
$\widetilde{D} (0;0,0)$ develops a pole. This may happen only
when $\widetilde{D}_{\rm sc}^{(0)} (0,0)$ becomes large enough 
at low temperatures. In the superconducting region
of the phase diagram, the correlations grow large in the limit
of vanishing temperature, as shown in Fig. \ref{corre}. The
only limitation to develop a real divergence is placed by the
finite length of the system, as we discuss later on.

The parameter that plays the major role in
setting the values of the transition temperature is the
weight $\lambda_2 (0)$ for pair hopping at zero
transverse momentum. One can look, for instance, for the
points of the phase diagram with transition temperature
$T_c = 10^{-3} E_c$ . This value corresponds to a temperature 
of $\approx 1 \; {\rm K}$ for $E_c = 0.1 \; {\rm eV}$. 
Taking the intermediate value $\lambda_2 (0) = 0.025$, the 
points form the boundary represented by the dashed line
in Fig. \ref{scsdw}. The curve has the same shape that the 
boundary of the superconducting phase determined from the 
expression (\ref{asd}) of the anomalous dimension. However, 
we see that the region with $T_c > 10^{-3} E_c$ above the 
dashed line is sensibly smaller than that of the whole
superconducting phase.

\begin{figure}
\begin{center}
\mbox{\epsfxsize 8cm \epsfbox{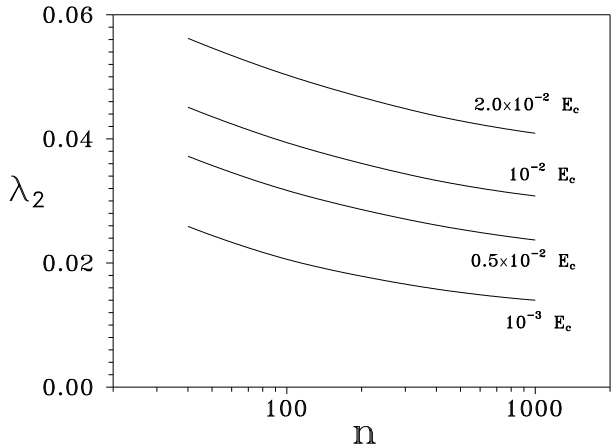}}
\end{center}
\caption{Contour lines for the critical temperature $T_c$ 
depending on the pair-hopping parameter $\lambda_2 (0)$ and 
the number $n$ of metallic nanotubes.}                                            
\label{lambdan}                                  
\end{figure}

We have represented in Fig. \ref{lambdan} the contour lines
for different critical temperatures in the space of the pair 
hopping parameter $\lambda_2 (0)$ and the number $n$ of metallic 
nanotubes, fixing the coupling of the attractive interaction
at $4|g|/\pi v_F = 0.75$. We observe that a slight change in the
value of the tunneling amplitude leads to a significant
increase in the transition temperature. Considering ropes with
larger content of metallic nanotubes may also help to enhance
$T_c$, although the figure shows that very high values of $n$
have to be reached to find a sensible variation.

\begin{figure}
\begin{center}
\mbox{\epsfxsize 8cm \epsfbox{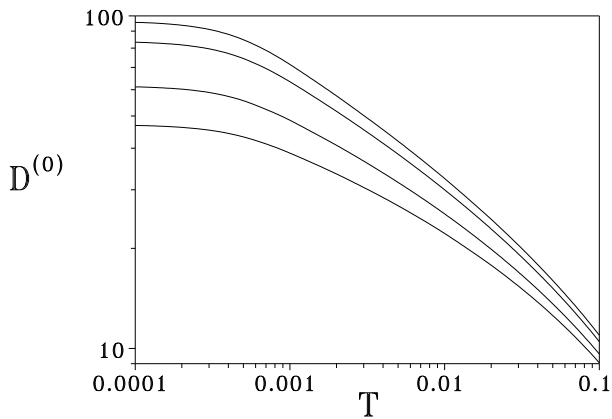}}
\end{center}
\caption{Logarithmic plot of $\widetilde{D}^{(0)}_{{\rm sc}}$ 
at zero frequency and momentum versus $T/E_c$, for 
$2 e^2/(\pi^2 v_F) = 1.0$ and $4|g|/(\pi v_F) = 0.75$. From 
top to bottom, the curves correspond to different numbers
of metallic nanotubes $n = 1000, 400, 100$ and 40 .}
\label{nlarge}
\end{figure}

Finally, we remark that the finite length of the nanotubes 
imposes a limit on the strength of the correlations. 
The existence of a length scale $L$ spoils the scaling of
the model and, therefore, the approximate power-law behavior 
of the correlators. Thus, the divergence of the propagators at
zero frequency and momentum is cut off in practice at a 
temperature scale which is between $v_F /L$ and one order of
magnitude below that value. This effect is illustrated in 
Fig. \ref{nlarge}, which displays the behavior of 
$\widetilde{D}^{(0)}_{{\rm sc}} (0,0)$ for ropes with different 
number of metallic nanotubes $n = 40, 100, 400, 1000$
and finite length $L = 1000/k_c$.

The constraint on the superconducting correlations due to the
finite-size scaling may have been observed in the experiments
reported in Ref. \onlinecite{koc}. It is shown there that
the large drop in the resistance is present in two samples
with respective lengths of $1.6 \; {\rm \mu m}$ and 
$1 \; {\rm \mu m}$. The effect is absent in a third sample 
which is $0.3 \; {\rm \mu m}$ long. 
Assuming that the typical transition temperature for these
samples is around $T_c \sim 0.5 \; {\rm K}$, the relative small 
length of the third sample would explain the absence of 
superconductivity. The minimum transition temperature that
could be supported in that case is of the order of
$T_c \sim 0.1 \; v_F/L \sim 10^{-4} \; {\rm eV }\sim 1 \; 
{\rm K}$. The same argument provides a definite check
of the model elaborated in this paper since, according to it, 
a transition with critical temperature $T_c = 0.1 \; {\rm K}$, 
for instance, should not be present in the samples below 
$1 \; {\rm \mu m}$. In general, it has to be true that 
any sample with length $L$ cannot have a transition at temperatures
lower than $0.1 v_F/L$, as it is observed in the samples 
considered in Ref. \onlinecite{koc}.

\section{Discussion}

In this paper we have shown that superconductivity is
a plausible effect in the ropes of nanotubes. The ropes with
greater number of metallic nanotubes have in general larger
superconducting correlations. We have seen that the long-range
Coulomb interaction is reduced very effectively in 
ropes with 100 metallic nanotubes or more. On the other hand,
the coupling to the elastic modes within each nanotube provides
the attractive interaction leading to the electronic pairing.
We have shown that the low-momentum optical phonons have suitable
properties to balance the effect of the repulsive Coulomb
interaction.

We have dealt with a model providing an exact description of
the competition between the Coulomb interaction and the effective 
attractive interaction in interbranch and intrabranch processes.
Proceeding in this way, we have disregarded the effect of
other processes coming from phonon exchange. These contribute to 
the backscattering and Umklapp couplings $g_1^{(1)}$,
$g_1^{(3)}$, $g_2^{(1)}$ and $g_2^{(3)}$. Only the first and the
third kind of processes play a role in the development of
the superconducting correlations, as analyzed in the Appendix
B. These coupling are marginally relevant in the renormalization
group sense but, as we have already remarked in the paper, the
effects derived from renormalization group scaling turn out to 
be very soft down to the energies where to look for a 
superconducting transition.

The significance of the backscattering processes is found in
that they determine the symmetry of the order parameter
whenever the system becomes superconducting. According to the
analysis in Section II, both $g_1^{(1)}$ and $g_2^{(1)}$
correspond to an effective attractive interaction. Then, from
inspection of the different order parameters considered in
the Appendix B, we conclude that singlet pairing is enhanced
by the backscattering interactions. We observe also
that the $s$-wave symmetry with positive amplitude in all the 
branches is favored over more exotic possibilities like the 
$d$-wave symmetry of the order parameter.

In our description of the mechanism of superconductivity, we 
have taken into account the fact that the ropes are made of 
nanotubes with a random distribution of helicities. This 
prevents the development of any ordered charge or spin structure 
from the coupling of the nanotubes. Moreover, this kind of 
disorder implies a strong suppression of the single-particle 
hopping and that the relevant tunneling process is given by 
the hopping of Cooper pairs between neighboring nanotubes. 
The amplitude for that process is in general very small, 
which explains the relatively low transition temperatures 
measured experimentally.

In any event, the effect of pair hopping has the most direct 
influence on the transition to the superconducting state.
The disorder present in the samples used in the experiments 
has to be very high, and its effect should be studied on more
quantitative grounds to have an estimate of the maximum
transition temperature reachable in the ropes. In this sense,
good prospects should exist to increase the transition 
temperatures, either by suitable refinement of the experimental
samples or by intercalation or modification of the internal
structure of the ropes.

Fruitful discussions with F. Guinea and A. Kasumov are
gratefully acknowledged. This work has been partly
supported by CICyT (Spain) and CAM (Madrid, Spain)
through grants PB96/0875 and 07N/0045/98.

\appendix

\section{Selection rules for the coupling to optical phonons}

The different symmetry of the modes with bonding and  
antibonding character leads to definite relations between the
various electron-phonon couplings. This can be observed in the 
results of Ref. \onlinecite{jdd}, where the full expressions of 
the couplings to the acoustic phonons in armchair nanotubes
have been obtained. In
this appendix we exploit the mentioned symmetry to get the 
corresponding relations in the case of the optical phonons.

The use of the tight-binding approximation is appropriate for
the carbon nanotubes\cite{pietro}. The electron-phonon coupling 
can be written in terms of the polarization vector 
$\mbox{\boldmath $\epsilon$}_s (k)$ depending on site $s$ and 
the amplitudes of the incoming and outgoing modes 
$u_s^{(p)} (k)$ and $u_{s'}^{(p')} (k')$ in 
the respective subbands $p$ and $p'$, taking the form
\begin{eqnarray}
g_{p,p'} (k,k')  & = &  \frac{1}{( \mu \; \omega_{k-k'} )^{1/2} }
  \sum_{\langle s , s' \rangle } 
    u_s^{(p) *} (k)   u_{s'}^{(p')} (k')   \nonumber   \\ 
   &   &     ( \mbox{\boldmath $\epsilon$}_s (k-k') - 
        \mbox{\boldmath $\epsilon$}_{s'} (k-k') ) 
      \mbox{\boldmath $\cdot \nabla$}   J (s,s')
\label{tight}
\end{eqnarray}
In the above expression, the sum is restricted to
nearest-neighbors of the atoms in the unit cell of the nanotube,
$J (s,s')$ is the matrix element of the atomic potential 
connecting orbitals at $s$ and $s'$, $\omega_k$ is the phonon
frequency and $\mu $ is the mass per unit length.

\begin{figure}
\begin{center}
\mbox{\epsfxsize 7cm \epsfbox{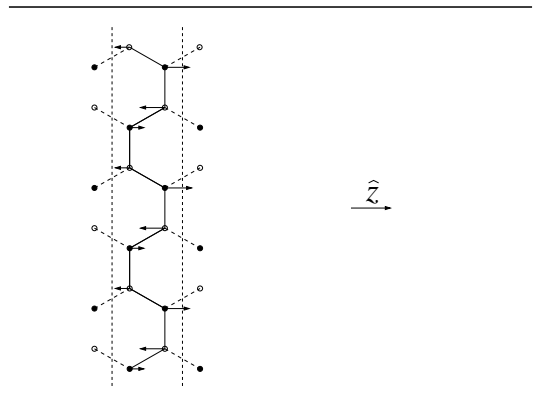}}
\end{center}
\caption{Scheme of the unit cell of an armchair nanotube, 
represented between the vertical dashed lines. The arrows
correspond to the displacements for a longitudinal optical
phonon.}
\label{long}
\end{figure}

Let us consider first the case of optical phonons with 
longitudinal polarization. The polarization vector depends only
on the distance $z$ measured along the nanotube, but it has
opposite amplitudes in the two sublattices shown in 
Fig. \ref{long} so that
\begin{eqnarray}
\mbox{\boldmath $\epsilon$}_s (k)  & =  & \hat{z}
 \exp (ikz_s)  \;\;\;\;\;\;\;\;\; {\rm for \;\; black \;\; points}  \\
\mbox{\boldmath $\epsilon$}_s (k)  & =  &  - \hat{z}
 \exp (ikz_s)  \;\;\;\;\;\;\;\;\; {\rm for \;\; white \;\; points}
\end{eqnarray}
$\hat{z}$ being the unit vector in the axis direction.

It is not difficult to see that, if the modes belong to the same 
subband $p = p'$, the different terms in the sum of the 
expression (\ref{tight}) cancel among themselves. Then, we have
that in the case of longitudinal optical phonons
\begin{equation}
g_{1,1} (k,k') = g_{2,2} (k,k') = 0
\label{long1}
\end{equation}
When the incoming and outgoing electron modes are in different 
subbands, the electron-phonon coupling does not vanish. Using the 
fact that the amplitude of the modes with antibonding character 
changes its sign under the exchange of the two sublattices, 
we obtain the result 
\begin{equation}
g_{1,2} (k,k') = - g_{2,1} (k,k')
\label{long2}
\end{equation}

\begin{figure}
\begin{center}
\mbox{\epsfxsize 7cm \epsfbox{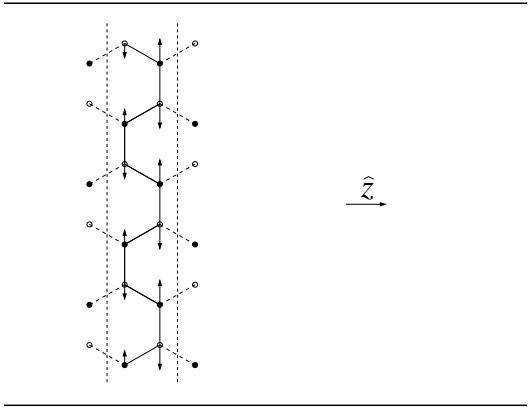}}
\end{center}
\caption{Same scheme as in Fig. \ref{long}, but representing
the displacements for a transverse optical phonon.}
\label{trans}
\end{figure}

Moving now to the case of transverse optical phonons, we have
a picture like that shown in Fig. \ref{trans}. The polarization 
vector is
\begin{eqnarray}
\mbox{\boldmath $\epsilon$}_s (k)  &  =  &  \hat{\theta} 
 \exp (ikz_s) \;\;\;\;\;\;\;\;\;   {\rm for \;\; black \;\; points}  \\   
\mbox{\boldmath $\epsilon$}_s (k)  &  =  &  - \hat{\theta} 
 \exp (ikz_s) \;\;\;\;\;\;\;\;\;   {\rm for \;\; white \;\; points}
\end{eqnarray}
$\hat{\theta} $ being a unit vector tangential to the nanotube. 
When the incoming and outgoing electron modes belong to the
same subband, we have now 
\begin{equation}
g_{1,1} (k,k') = - g_{2,2} (k,k')
\label{trans1}
\end{equation}
If the modes are in different subbands, the terms in the 
expression (\ref{tight}) cancel out in pairs and we have
\begin{equation}
g_{1,2} (k,k') =  g_{2,1} (k,k') = 0
\label{trans2}
\end{equation}

We observe that the situation is reversed with respect to the 
case of the acoustic phonons, in which the relations 
(\ref{trans1}) and (\ref{trans2}) apply to the longitudinal branch
while (\ref{long1}) and (\ref{long2}) hold for the transverse
branch.

\section{Pairing symmetry of the superconducting correlations}

There are several response functions that give a measure 
of the superconducting correlations in the nanotubes, each 
of them corresponding to a different symmetry of the pair
wavefunction. They have the general form
\begin{equation}
R(x,t) = \langle {\cal O} (x,t) 
      {\cal O}^{\dagger} (0,0) \rangle
\end{equation}
In the case of singlet and triplet pairing, the operator
${\cal O} (x,t)$ corresponds respectively to the upper $-$
sign and the lower $+$ of the expression
\begin{eqnarray}
{\cal O} (x,t)  & = & \Psi_{++\uparrow}  
       \Psi_{--\downarrow}  \mp  \Psi_{++\downarrow}  
         \Psi_{--\uparrow}       \nonumber \\   
  &  & + \Psi_{-+\uparrow}   \Psi_{+-\downarrow} 
       \mp  \Psi_{-+\downarrow} \Psi_{+-\uparrow} 
\end{eqnarray}
There are also more exotic possibilities like the $d$-wave 
symmetry of the Cooper pairs, corresponding to
\begin{eqnarray}
{\cal O} (x,t)  & = & \Psi_{++\uparrow} \Psi_{--\downarrow}
   -  \Psi_{++\downarrow}  \Psi_{--\uparrow}  \nonumber \\
  &  & - ( \Psi_{-+\uparrow} \Psi_{+-\downarrow}
       -  \Psi_{-+\downarrow}  \Psi_{+-\uparrow} )
\end{eqnarray}

\begin{figure}
\begin{center}
\mbox{\epsfxsize 7cm \epsfbox{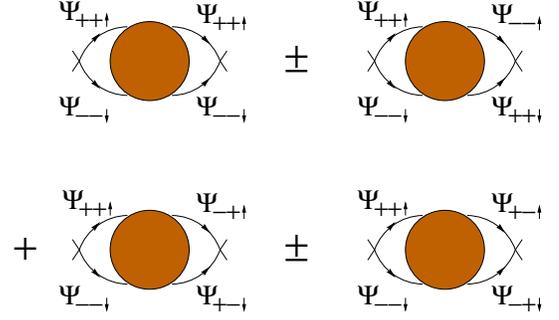}}
\end{center}
\caption{Contributions to the response functions for singlet and 
triplet pairing, where the filled circles represent the different
renormalized vertices.}
\label{pairs}
\end{figure}

The symmetry of the superconducting correlations 
can be obtained by using a diagrammatic
analysis of the different response functions. In the case of
singlet and triplet pairing, the contributions to the 
correlator have the structure depicted in Fig. \ref{pairs},
with respective upper and lower signs. Then, the enhancement
of the singlet pairing response function is driven by the
combination of renormalized couplings $g_2^{(2)} + g_1^{(1)}
+ g_2^{(1)} + g_1^{(2)}$, while in the case of triplet 
pairing the combination is $g_2^{(2)} - g_1^{(1)} + g_2^{(1)} 
- g_1^{(2)}$.

\begin{figure}
\begin{center}
\mbox{\epsfxsize 7cm \epsfbox{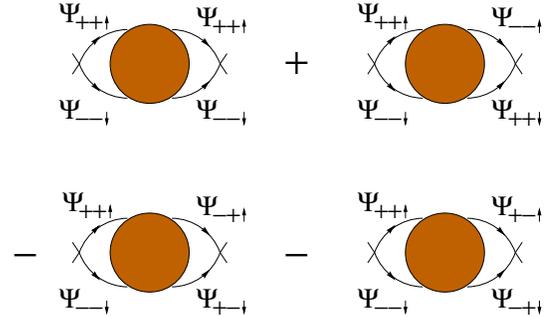}}
\end{center}
\caption{Contributions to the response function for $d$-wave
pairing, with the same representation as in Fig. \ref{pairs}.}
\label{paird}
\end{figure}

For the response function with $d$-wave symmetry of the 
Cooper pairs, the structure of the contributions is represented
in Fig. \ref{paird}. The enhancement is controlled now by the 
combination of couplings $g_2^{(2)} + g_1^{(1)} - g_2^{(1)}
- g_1^{(2)}$.

\begin{references}
\bibitem{foot}
A phase of this kind has been also found in Ref. 
\onlinecite{loss} in the study of the 1D electron system 
coupled to acoustic phonons.
\bibitem{priv}
This is the case of the superconducting ropes
of Ref. \onlinecite{koc}, although to meet such
experimental condition one has to find the appropriate
sample out of a large number of them.
\end{references}
\end{document}